\documentstyle[12pt]{article}
\begin{document}
\vspace{0.5cm}
\centerline{Dedicated to the 70th birthday of Mikio Sato}
\vspace{0.3cm}
\centerline{\Large S.Novikov\footnote{S.P.Novikov,University of Maryland,
College Park, MD, 20742-2431 and Landau Institute for Theoretical Physics,
Moscow 117940, Kosygina 2, e-mails novikov@ipst.umd.edu
 and novikov@landau.ac.ru,
fax 301-3149363, phone 301-4054836(o)}}
\vspace{0.3cm}
\centerline{\Large Discrete Schrodinger Operators and Topology\footnote{
This work was supported by the NSF Grant DMS9704613}}
\vspace{0.2cm}
{\bf Introduction}.

During the last 3 years the present
 author made a series of works \cite{NV1,NV2,
N1,N2,N3,NT,ND,DN,N4} dedicated to the study of the unusual spectral
properties of  low-dimensional continuous
and discrete (difference) Schrodinger Operators.
 Some of these works were done in collaboration with
A.Veselov, I.Taimanov and I.Dynnikov. First, let me briefly describe
the list of
problems discussed in these works.

1.Euler-Darboux-Backlund (EDB)-Transformations as 
 nonstandard spectral symmetries for
the 1-dimensional Schrodinger Operators and its discrete analogs
 on the lattice $Z$. Problem of cyclic chains, its solutions for
 the special cases. Exactly solvable spectral problems for some
 operators. EDB Transformations for the nonstationary
 Schrodinger Equation and the problem of cyclic chains (see \cite{W,S1,S2,SV,
 SVZ,NV2,NT,M}).

 2.Laplace Transformations for the 2D stationary Schrodinger operators
in the double-periodic magnetic field and potential,
 acting on the space of eigenfunctions of one energy level.
  Problems of cyclic,
 semicyclic and quasicyclic chains.
 The possibility to have two exactly solvable highly degenerate
  energy levels as a maximal possible
 solvability
  for the spectral theory in the Hilbert space
 $L_2(R^2)$ (except  the Landau case in  constant   magnetic field
 and trivial potential)? Discretization of the Laplace transformations:
 square lattice is compatible with hyperbolic equations; equilateral
 triangle lattice is compatible with elliptic selfadjoint operators.
 Exactly solvable operators. See \cite{NV1,NV2,N1,N2,ND}.

 3.The second order operators on  simplicial complexes. Factorizations
 and Laplace Transformations. The cases of  2-manifolds with 2-colored
 triangulation and
   multidimensional
 equilateral lattices. Zero modes problem. 
 First order equations in the simplicial complexes and
 nonstandard discretization
 of connections, combinatorial curvature. See \cite{ND,DN}.

 4.Schrodinger Operators on simplicial complexes. The combinatorial analog
 of  Wronskians--the {\bf  Symplectic Wronskians or SWronskians}
  in our terminology; their topological properties. Special case of graphs with
 finite number of tails. Scattering Theory and Symplectic Geometry.
 See \cite{N3,N4}.

 This work is a direct continuation of \cite{N3,N4}  (the idea 
 was quoted in these papers communicated to the present author by
 I.Gelfand in 1971 as a reaction to the authors works \cite{N5},
 where Symplectic Algebra was used for the needs of Differential
 Topology). We extend here
 the definition and topological properties of the Wronskians (Symplectic
 Wronskians or SWronskians) to the
 broad class of operators on the simplicial complexes.

\vspace{0.3cm}
\centerline{\Large 1.Finite order selfadjoint combinatorial operators.}
\centerline{\Large Symplectic Wronskians and Topology.}
\vspace{0.2cm}
Let us consider any locally finite simplicial complex $K$
 where any simplex belongs to the finite number of simplices only.

 By definition, the {\bf Distance between two simplices} of any dimensions
 $d(\sigma,\sigma')$ is equal to zero if and only if they coinside.
 It is equal to $1/2$ if and only if one of them belongs to the boundary
 of the other one. It is  equal to $s/2$ if $s$ is such
 a  minimal number $s$ that there exists
 a {\bf simplicial path}, i.e. sequence of simplices
  $$\sigma=\sigma_0,\sigma_1,\ldots,\sigma_s=\sigma'$$
  \noindent with
 $d(\sigma_j,\sigma_{j+1})=1/2,j=0,1,\ldots,s-1$.

 {\bf The Operators $L$ of the order less or equal to $k$}
  we define by the formula
 \begin{eqnarray}
L\psi(\sigma)=\sum_{\sigma'}b_{\sigma :\sigma'}\psi(\sigma')\\
 d(\sigma,\sigma')\leq k/2\nonumber
 \end{eqnarray}

Here $\psi(\sigma)$ belongs to some space of  real or complex
scalar-valued or 
vector-valued functions on the set of simplices.

 The operator    $L$
is {\bf Symmetric} iff $b^{ij}_{\sigma :\sigma'}=\bar{b}^{ji}_{\sigma' :\sigma}$.

The operator is {\bf Real} iff all coefficients $b$ are real.

For the {\bf Second Order Operators} we have exactly
 $d(\sigma,\sigma')\leq 1$ for the nontrivial coefficients
 $b_{\sigma :\sigma'}$. Some nontrivial coefficients should be such that
 $d=1$ exactly. For the {\bf Homogeneous Operators} of some order $k$
 we have $d=k/2$ for all nontrivial coefficients.

 In the previous works we
 restricted our attention to the case where all nonzero
 coefficients $b_{\sigma_1^p :\sigma_2^s}$
 are concentrated on the simplices of some
 specific dimensions $p,s$. In this case the
 operator maps the space of functions (or vector--functions)
 on the set of $p$-simplices  into the space of functions  (vector--functions)
 on the set of $s$-simplices:
 \begin{eqnarray}
 L:C^p\rightarrow C^s
 \end{eqnarray}

\noindent  We call them {\bf the operators
  of the type $p,s$  }.
  The most interesting classes are as follows:

  1.The second order selfadjoint (i.e. Schrodinger) Operators for $p=s$.

  2.The first order operators of the type $p,s$. Especially interesting
  is the case $p\pm 1=s$, but other cases also appeared before (see \cite{ND}).

\noindent    The symmetric (hermitian) matrix-function
 $V(\sigma)=b_{\sigma :\sigma}$ will be called
 {\bf Potential}.

 \noindent  Let us consider the real operators acting on
  the $l$-component vector-valued functions $\psi(\sigma)$,
    where $\sigma\in K,\psi=(\psi^j)\in C^l,j=1,\ldots,l$, and $\sigma$
  is a simplex of any dimension. The operator $L$ acts on the space
  $C^*=\oplus_p C^p(K)$ where summation is extended to all dimensions
  (it is a full set of vector-valued cochains).

  In the standard way we define  a ''baricentrical'' subdivision
   of the simplicial
   complex  $K$. We put new vertices (0-simplices) in the centers of all
   original simplices of all dimensions $k\geq 0$. After that new edges
   connect  the centrum of every simplex with all new vertices
   located on its boundary. The new $k$-simplices of any dimension
   are exactly the cones looking from the centers of the old simplices
   into the  new $k-1$-simplices already constructed by the induction
   on the boundary. We denote the baricentrical
   subdivision of the simplicial complex $K$ by $K'$.

   Consider now any real symmetric operator $L:C(K)\rightarrow C(K)$
   of the  order $k$,
   acting on the space of all vector-valued cochains.

   {\bf
    Any such operator can be
   treated as an Operator $L'=L$ of the type (0,0) acting  on the zero-dimensional
   cochains in the baricentrical subdivision $K'$   }:
   $$L':C^0(K')\rightarrow C^0(K')$$
\noindent    Take any pair of  solutions for the
   equation
   $$L'\psi=\lambda\psi, L'\phi=\lambda\phi$$
   \noindent  {\bf For every pair of vertices $\sigma\sigma'\in K'$ fix
    a unique naturally oriented path $l(\sigma,\sigma')$
    (i.e. 1-chain $[l]$) such that
     $\partial [l(\sigma,\sigma')]=\sigma'-\sigma$.
      Let for conveniency this path be the one
   of the minimal lengh}. For the cases $d(\sigma\sigma')\leq 1$
   such a path is unique. 
   It is always unique for any pair of vertices in every
   simply-connected Graph (tree). It is also unique for the pairs of vertices
   if the distance between them is small enough:
   $d(\sigma,\sigma')< 1/2 d_0$
   \noindent where $d_0$ is a size of the smallest 1-cycle, $d_0/2$ is
   the number of edges in it.
   \newtheorem{deff}{Definition}
   \begin{deff}  The Symplectic Wronskian (SWronskian)
   for the pair of solutions for the operator
   $L'$ of the type $(0,0)$ in any simplicial complex $K'$  is
   a one-dimensional (possibly infinite)
   simplicial chain $W(\psi,\phi)$ in the complex $K'$
   defined by the formulas below:
   \end{deff}

   \begin{eqnarray}
   W(\psi,\phi)=\sum_{\sigma\sigma'}W_{\sigma\sigma'}(\psi,\phi)\\
   W_{\sigma\sigma'}=\sum_{ij}b^{ii}_{\sigma:\sigma'}
   \{\psi^i(\sigma)\phi^j(\sigma')-
   \phi^j(\sigma')\psi^i(\sigma)\}[l(\sigma\sigma')]\nonumber
    \end{eqnarray}
\noindent
For the locally finite complex and finite order operator $L'$
this sum makes
 sense as an infinite chain in this complex. {\bf We consider
 the operators $L$ acting
 on the
simplices of any dimension in the complex $K$  as the operators $L'$
acting on the vertices of the baricentrical subdivision $K'$.
 Therefore we defined the SWronskians
 for all selfadjoint real operators
of any finite order $k\geq 1$ acting on the spaces of vector--valued functions
on the set of simplices of all dimensions.}

   \newtheorem{th}{Theorem}
   \begin{th}  The Symplectic Wronskian (SWronskian)
   defined above as a C-valued
   finite or infinite 1-chain in $K'$
   is in fact an open cycle, i.e. $\partial W=0$. This cycle is a bilinear
   skew-symmetric functional of the pair of solutions for the equation
   $L\psi=\lambda\psi,L\phi=\lambda\phi$.
   \end{th}

\newtheorem{rem}{Remark}
\begin{rem}Let any  solution $L\psi=\lambda\psi$ be given describing
in the sense of  Quantum Mechanics the stationary state of  electron,
living in the simplicial complex $K$ with Hamiltonian $L$ and energy $\lambda$.
 This state defines a {\bf
Quantum Current} $J(\psi)=W(\psi,\bar{\psi})$ along the arcs in $K'$
satisfying to the {\bf Kirchhof Law} in every vertex.
\end{rem}

Proof of the theorem.

Consider the expression $\sum_i\phi^i(\sigma)(L\psi)^i(\sigma)-
\psi^i(L\phi)^i(\sigma)$ for the pair of
vector-functions. We can easily see that all zero order terms
containing $b_{\sigma:\sigma}$ disappear from this expression
 obviously for the
real operators.

\noindent For any vertex $\sigma$ of the complex $K'$ we should
consider all 1-simplices of $K'$ meeting each other
 in the vertex  $\sigma$.
By definition of the Wronskian, we have
$$(\partial W)_{\sigma}=\sum_{\sigma''}W_{\sigma\sigma''}$$

\noindent where either $\sigma$ is a nontrivial face of $\sigma''$
 or vise versa, i.e. $d(\sigma,\sigma'')=1/2$.
 At the same time,
  $(L\psi)_{\sigma}=\sum_{\sigma'}b_{\sigma:\sigma'}\psi(\sigma'$.
  Canceling from the expression
 $\phi(\sigma)L\psi(\sigma)-
 \psi(\sigma)L\phi(\sigma)$ all zero order terms, we group others 
   in such a way that our expression looks as a
  sum of the ''elementary Wronskians''
   $\sum_{\sigma''}W(\psi,\phi)_{\sigma\sigma'}$.

 After that we memorize that $\psi,\phi$ are in fact the solutions
 for the equation $L\phi=\lambda\phi,L\psi=\lambda\psi$, so our expresssion
 is equal to zero.
 Theorem is proved.
 \newtheorem{cor}{Corollary}
 \begin{cor}
 Let $K$ is a Graph, i.e. $dimK=1$. For any second order operator $L$
 acting on the full space of vector-valued cochains $C=C^0\oplus C^1$
  and any pair of solutions
 $\psi,\phi$ for the spectral problem,  their SWronskian
 is an open cycle (i.e. open homology class) in the same Graph
 $$W(\phi,\psi)\in H_1^{open}(K,C)$$
 \end{cor}

\noindent Proof. For  graphs every simplicial 1-cycle in $K'$ is in fact a
simplicial 1-cycle in
  the original graph $K$.
  \begin{rem}
  Let us point out that we already proved and used this observation for the
  scattering theory on the graphs--see \cite{N3,N4}. However, in these
  works we considered   strictly homogeneous
  second order operators only, acting
  on the spaces of vertices $C^0\rightarrow C^0$
  or edges $C^1\rightarrow C^1$ separately. We also defined in  \cite{N4}
  the SWronskians for the higher order operators acting on the space of vertices
  and SWronskians for the strictly homogeneous second order operators on the
   simplices of every fixed dimension. 
Here we extend the class of admissible operators. In
particular we may work with
 operators $L:C^*\rightarrow C^*$
 mixing cochains of the different dimensions.

 {\bf All previous authors'
 definitions of the Wronskians as a symplectic (skew-symmetric bilinear)
 vector--valued 2-forms are the partial cases of this one.}
\end{rem}

\newtheorem{ex}{Example}

For any simplicial complex $K$ there is a famous
selfadjoint first order 
operator $L=d+d^*:C(K)\rightarrow C(K)$ where $d=\partial^*:C^k\rightarrow C^{k+1}$
and $d^*=\partial:C^k\rightarrow C^{k-1}$ for every value of the dimension $k$.
Its square is a direct sum or the Laplace -Beltrami Operators
$\Delta_k=dd^*+d^*d:C_k\rightarrow  C_k$. For the finite complexes
zero modes of the operators
$L,\Delta$ 
give certain ''Harmonic'' basis for the Homology (Cohomology) Groups
$H_k(K,R)$. Both these Operators are selfadjoint. They are the Euler-Lagrange
operators for the quadratic functionals:

\begin{eqnarray}
S_{\Delta}(\psi)=<\psi,\Delta\psi>=<d\psi,d\psi>+<d^*\psi,d^*\psi>\\
S_L(\psi)=<\psi,(d+d^*)\psi>\nonumber
\end{eqnarray}

In the elasticity theory for the isotropic media the linear combinations appear
$\lambda dd^*+\mu d^*$ acting on 1-forms, where $\lambda,\mu$
 are the Lame' parameters
(in the continuous case).

\begin{rem}For the zero modes of the Laplace-Beltrami
Operators $\Delta_k$ on the finite simplicial complexes
we can easily prove that their SWronskian is always identically
equal to zero.
\end{rem}

\begin{ex} In the works \cite{NV2,N2,ND,DN} factorizations and
Laplace Transformations were considered on the 2-colored (black and white)
triangulated
two-manifolds $M^2$ for the different classes of  Schrodinger Operators.
In the case of vertices we consider the operators $L:\psi(P)=\sum_{P'}b_{P:P'}\psi_{P'}$
where $P'$ is such that $d(P,P')=2$. These real selfadjoint
operators can be factorized
in the Laplace-type (''weak'') form  $L=QQ^t+V$ where V is a ''potential'', i.e.
multiplication by the real function, and $Q^t:C_0\rightarrow C_2^{black}$.
It means that this first order operator $Q$   maps functions on the set
of vertices into the functions on the set of the black triangles.
Such an operator is defined by the set of all coefficients $c_{P:T}$
where $T$ is a black triangle and $P$ is one of its vertices.
 So the simplicial
 complex $K$ in this case is
$M^2$ minus white triangles. It has the same vertices and edges as $M^2$,
but twice less number of triangles.  In the case $V=0$  the ground level
(if it is equal to zero),
can be found from the square integrable solutions for the
first order {\bf Triangle Equation}
 $Q^t\psi=0$. Especially interesting is the classical case
 of the equilateral lattice $Z^2$ considered as a triangulation of $R^2$.

 \end{ex}

 For the Graphs $K=\Gamma$ several examples were considered in the
 work \cite{N4}, especially for the graphs with finite a number of
 infinite tails. We shall come to this later in connection with
 the  Scattering Theory.

  Let us consider here the special case of
 the discretized line with vertices numerated by the even numbers
 $2n=\sigma_n^0,n\in Z$ and edges  numerated by the odd integers
 $2n+1=\sigma^1_n=[2n,2n+2],n\in Z$. So we have a lattice $Z'$ of the
 integers  as a complex $K'$. The operator $L$ in $K$
 determines the operator $L'$ in $K'$ as a $(0,0)$ type one:
 $$(L'\psi)^i(n)=\sum_{j,s}b^{ij}_{n,n+s}\psi^j(n+s),-k\leq s\leq k$$
 \noindent We choose a basis $C_m$ of the solutions
  $$C_{m;p}^i,i=1\ldots,l,
 m\in Z, p=-k+1,-k+2,\dots,k-1,k$$
  in the form:

 \begin{eqnarray}
 (C_{m;p}^i)^j(m+s)=\delta^{ij}\delta_{ps}
 \end{eqnarray}

Let us compute the Symplectic Wronskian form in this important case.
This form is a scalar-valued skew-symmetric bilinear form because
there is only one basic geometrical cycle, the oriented line itself.

 \begin{th}The Symplectic Wronskian form written in
 the basis $C_m$ of the solutions
 $C_{m;p}^i$ for any given integer $m$  admits
  two $k$-dimensional Lagrangian Planes $L_{\pm}$
 (i.e.  this form is equal to zero on these planes),
 with basises $C_{m;p}^i\in L_+$ for $p=-k+1,\ldots,0$ and
$C_{m;p}^i\in L_-$ for $p=1,\ldots,k$. For the SWronskian scalar product
between these two planes we have 
 $$W(C_{m;p}^i,C_{m;q}^j)=0,q-p>k,p\leq 0,q\geq 1$$

$$ W(C_{m;p}^i,C_{m;q}^j)=b^{ij}_{m+p,m+q},q-p\leq k$$

 In particular, this form is nondegenerate if and only if the Operator
 $L'$ has everythere  nondegenerate higher order terms
 $$\det b^{ij}_{n,n+k}\neq 0$$
  for every $n\in Z$.

  \end{th}

Proof.We can easily verify the form of this matrix from the definition
of the basis and SWronskian above. The matrix of SWronskians can be
considered naturally as a number-valued one because there is
only one canonical
open geometrical 1-cycle on the line $R=K=K'$ oriented in the direction of $n\rightarrow
+\infty$. We compute the value of SWronskians $W(\phi,\psi)$ for all
solutions from our basis on the 1-edge $[01]$. Therefore only those pairs of
vertices should be considered which contain the segment $[01]$.
Otherwise the elementary SWronskian for the pair of vertices would not
contain $[01]$. It means in particular that we may have a nonzero
SWronskian in our basis only between the subspaces $L_+$ and $L_-$.
 The value of the  spectral parameter $\lambda$
does not affect this matrix in the given basis. Let us point out that the
matrix $SW_m$ of SWronskians $W(L_+,L_-)$ in our basis is a block-triangle one,
where the matrices $b_{m+p;m+p+k},p=-k+1,\ldots,0$ are positioned along
the diagonal. We have zero SWronskians below this block-diagonal part.
Therefore the determinant of this matrix is a product
$$\det SW_m=\prod_{p=-k+1}^{p=0}(\det b^{ij}_{m+p,m+p+k})^2$$

This implies the nondegeneracy of the Symplectic form given by the
SWronskians. Therefore our theorem is proved.

The SWronskian form is equal to constant along the line according to the
 theorem 1. We have following

\begin{cor} The Evolution Map $T_{m,m+1}(\lambda)$
from the basis $C_m$ to the basis $C_{m+1}$, given by the equation
$L\psi=\lambda\psi$ is a Linear Symplectic Transformation.
\end{cor}

 This theorem, of course, is very clear. It is valid also for
 the nonlinear systems, as we shall see later. Its continuous analog
 has been known
 many years. However, even in the continuous  case there was some difficulcy
  in finding the canonically adjoint (''Darboux'') basis, following the
  ''Ostrogradski Transformation'' for the vector-valued higher order
  variational
  problems (B.Deconinck pointed this out  to me).
   We don't try to find a canonical basis,
  but nondegeneracy of the Symplectic form is very easy in our case.
  The nonlinearity is unimportant in these problems. Probably,
   no one considered
  the discrete variational problems in  classical mathematics.
  This business
  was used in the Theory of Solitons for the discrete linear second order
  systems in the theory of ''Toda Lattice'' and ''Discrete KdV'' since
  the works \cite{F,Mn}. It was
  started for the second order nonlinear systems
  in  \cite{V}. For the discrete systems of 
  higher order we do not know
  any literature. Our main idea is that for the graphs
   and simplicial complexes
  instead of
  line as a  time
  we  have a Symplectic form taking values in the linear
   space of the open one-cycles $Z_1^{open}(K',C)$.

   Consider now any Graph $\Gamma$ which is presented as a regular
   $Z$-covering over the finite Graph $\Gamma_1$ with free simplicial
   action of the monodromy group
   $Z$ generated by the map $F$ :
   $$P:\Gamma\rightarrow\Gamma_1, F:\Gamma\rightarrow\Gamma, FP=PF$$

   \begin{th} Any  Operator of the finite order in the Graph $\Gamma$
   with free $Z$-action and finite factor can be
   presented as a higher order vector-valued
   operator on the discretized line--lattice  $Z$. The operators with
    $Z$-invariant coefficients      (i.e. operators whose coefficients
    are coming from the Graph $\Gamma_1$)
will be presented as operators with constant coefficients
 on the discretized line.
   \end{th}

 We call this presentation a {\bf Direct Image} of the operator on the lattice $Z$.

   Proof. For the proof, we construct a  map $f:\Gamma\rightarrow
   R$ commuting wth free $Z$-action.     It certainly exists.
   First of all we choose ''initial vertices'' in one-to-one correspondence
    with vertices of the factor-space $\Gamma_1$. It is
    good to choose them in the
    ''Fundamental Domain'' of the minimal size for the group $Z$ in $\Gamma$,
    starting from any initial vertex. We map all
     these initial vertices into $0\in Z$. After that we map all other vertices
     following the group action. The continuation to the 1-skeleton of
     $\Gamma$ is easy: for any edge its boundary vertices already mapped
     into $R$. The linear continuation is unique. It might happen that
      image of the edge
     is an interval $n,n+k$ where $k>1$.
     Therefore the original edge should be subdivided
     in $k$ parts. After that we have a simplicial $Z$-invariant map.
     Any function on the vertices of the Graph $\Gamma$ can be naturally
     and tautologically presented as a vector-valued function $\psi_n$
     on the vertices
     of the discretized line $Z$ with the
     number of components of vector $\psi_n$
      numerated by the vertices from $f^{-1}(n)\in \Gamma$.
      After this presentation of the functional space, we can see that the same operator
      looks as an operator of finite order on the lattice $Z$. Theorem is proved.

\vspace{0.3cm}
{\Large 2.Scattering and Symplectic Geometry}
\vspace{0.2cm}

As it was pointed out in  paragraph 1, any combinatorial
Schrodinger Operator $L$ of the order $k$,
acting on the functions on the set of simplices of any dimension,
can be considered as an operator $L'$ of the order $2k$ acting on the
vertices of the baricentrical subdivision $K'$. Therefore it depends on the
1-skeleton of the complex $K'$ only. So we shall consider any
higher order real selfajoint operator $L$ acting on the vertices
 of the Graph $\Gamma$. For the Scattering Theory we need to consider the
  following picture:

  1.Our Graph $\Gamma$ has a finite number $N$ of ''tails''
   (i.e. subgraphs $z_j, j=1,\ldots,N$,
  isomorphic to the ''half'' of the special ''line-like''
  graphs $K_j$--graphs with free action of the group $Z$
  generated by the map $F_j$,
   and finite factor $K_j/Z$). In particular, the map $F_j$ is well defined
   far enough into the tail.
  After removal of the tails, what remains is a finite subgraph $\Gamma'$.

  2.All coefficients of the Operator $L$ rapidly enough tend to constants
  (i.e. $F_j$-independent)
   in every tail $z_j, j=1,\ldots,N$. So, far enough in every tail $z_j$
   we have an operator
   $L^{as}_j$ with asymptotically constant coefficients.
    The vertices in every tail $z_j$ are
   numerated by the positive integers $n\geq 0$ and by the finite number of
   vertices of the factor-graph $K_j/Z$. The map $f_j:K_j\rightarrow R$
   is given of the tail
   into the discretized line (see the end of  paragraph 1), commuting
    with the
   action of $Z$. Therefore our operator far enough in the tail is presented as
   an operator on the discretized line.

\begin{deff} The solution $\psi$ for the equation $L\psi=\lambda\psi$
belongs to the {\bf spectrum}
 of the operator $L$ in the Hilbert Space $L_2(\Gamma)$
of the square integrable complex vector-functions on the Graph iff
its growth in the tails is less than exponential, i.e. there exists
a number $s$ such that $|\psi_{j,n}|<n^s$ for all tails $z_j$ and
$n\rightarrow+\infty,n\in Z$.

The solution $\psi$ belongs to the
{\bf discrete spectrum} of the operator $L$ iff
$\sum_{\sigma\in \Gamma} |\psi(\sigma)|^2<\infty$. In particular,
it is sufficient to require that $\sum_{j,n} |\psi_{j,n}|^2<\infty$
for all tails $z_j$. The eigenfunction is {\bf singular} iff it
 is equal to zero in all tails.
\end{deff}

For the operators $L^{as}_j$ with constant coefficients we describe
all solutions through the one symplectic matrix $T_j=T^{as}_{j;n,n+1}$ defined
in  paragraph 1.  This matrix expresses the basis $C_n$ through the basis
$C_{n+1}$ in the neighboring point. We obviously have the discrete evolution
for any $n<m$ where $T^{as}_{j;n,m}$ depends on  $m-n$ only for the operators with
constant coefficients:
$$T^{as}_{j;m,n}=T^{as}_{j;n,n+1}T^{as}_{j;n+1,n+2}\ldots
 T^{as}_{j;m-1,m}=T_j^{m-n}$$
\noindent Therefore the eigenvalues $\mu_{j;r}(\lambda)$
of the matrices $T_j(\lambda)$
 in the tail $z_j$ determine the asymptotic properties
of the eigenfunctions in the tails
 except of the ''singular part'' nonvisible from the tails. The structure
 of operator $L$ inside of the graph leads to the algebraic relations
  between the tails.

According to the modern textbook literature (see \cite{AG},paragraph 4),
the eigenvalues of the {\bf generic} real one-parametric $\lambda$-family of
the symplectic matrices $T_j(\lambda)$ are crossing in the isolated points
$\lambda_*\in R$
 the so-called ''codimension 1
degeneracies'':

Path  1. It may have a pair of Jordan blocks of  length 2
corresponding to the pair of real eigenvalues $\mu_1=
(\mu_2)^{-1}\neq \pm 1$;

Path  2. It may have a pair of Jordan blocks of  length 2 corresponding
 to the pair of unimodular complex eigenvalues $\mu_1=\bar{\mu}_2\neq\pm 1$;

Path 3. It may have  a unique Jordan block of  length 2 corresponding to
the eigenvalue $\mu_1=\pm 1$.

 All other eigenvalues remain
simple during these processes.

Let us remind that the eigenvalues of any symplectic matrix are invariant
under the complex conjugation $\mu\rightarrow \bar{\mu}$
and inversion $\mu\rightarrow \mu^{-1}$.

 Therefore the symplectic  $2M\times 2M$-matrix $T_j(\lambda)$
has in the generic point $\lambda_*$ of the $\lambda$-line:

I. some number $s$ of the nonmultiple unimodular eigenvalues $|\mu_i|=1$
in the upper halfplane $Im(\mu_i)>0$ (and their complex adjoint),
not equal to
the $\pm 1$.

II. $2p$ nonmultiple nonreal eigenvalues inside of the unit circle $|\mu_|<1$
and the same number outside of the unit circle.

III. $q$ nonmultiple
real eigevalues inside of the unit circle and the same number
outside of the unit circle.

So we have  the total dimension $2M$ equal to $2M=2s+4p+2q$.

\begin{ex}a) For the second order scalar operators or first order
2-vector-valued
operators we have $2M=2$. Therefore $p=0$. We have either $s=1$ or $q=1$.
In the isolated points $\lambda_*$ our generic family is passing through
the Jordan block of  length 2 with eigenvalue $\pm 1$.

b)For the case $2M=4$ we may have for $p,q,s$ the following possibilities:
$$(p,q,s)=(1,0,0);(p,q,s)=(0,2,0);(p,q,s)=(0,1,1);(p,q,s)=(0,0,2)$$
\end{ex}

In the isolated points this family may pass through the Jordan blocks
of the types and multiplicities described above corresponding to the
multiple
eigenvalues on the unit circle, on the real line or in the special
points $\pm 1$. For the Paths 1-3 we have:

Path 1  transforms $(p,q,s)$ into $(p+1,q-2,s)$ or vice-versa;
two real eigenvalues collide with each other inside of the unit circle and
transform into the complex adjoint pair or inverse process.
{\bf This process is unimportant for the Scattering Theory}.
Only Paths 2 and 3 where the number $s$ changes
 are important for the Spectral
theory in the Hilbert Space $L_2(\Gamma)$.

 We do not see any spectral singularity
in the point $\lambda_*$ critical for  Path 1.

Path 2 transforms $(p,q,s)$ into   $(p+1,q,s-2)$ or vice-versa;
two unimodular eigenvalues collide with each other in the upper part of the
circle and transform into the pair inside and outside of this circle
or inverse process. The structure of the Continuum Spectrum may be
drastically changed
in this point.

Path 3 transforms $(p,q,s)$ into $(p,q+1,s-1)$ or vice-versa;
two real eigenvalues collide with each other in the point $\pm 1$
 and transform
into the unimodular complex adjoint pair or inverse process.
This path also  changes the structure of the spectrum.

{\bf We assume that the family $T_j(\lambda)$ is generic in the sense described
here, for all tails $z_j$}. Let us point out that our families
 $T_j(\lambda)$ have very special  $\lambda$-dependence. Therefore
 this assumption in fact should be verified  in the future for the generic
 operators with constant coefficients. It is certainly true for the second
 order scalar operators--it is almost obvious and was used
  in the literature many times. For the higher order operators
  and matrix operators
  we shall return to this in the later publications.

\begin{deff} The solution $\psi$ for the equation $L\psi=\lambda\psi$
is a point of the {\bf regular discrete spectrum} iff  in every
tail $z_j$ it belongs asymptotically to the linear span of the
eigenspaces corresponding to the eigenvalues of $T_j(\lambda)$ inside
of the unit circle for every $j$.
\end{deff}

Let us consider now the special important case:

 All asymptotic
operators $L^{as}_j$ coinside with each other $L^{as}_j=L^{as}$;

For the asymptotic operator $L^{as}$ there is
a nonempty interval $[\lambda_0,\lambda_1]\subset R$ on the $\lambda$-line
such that for all $\lambda\in [\lambda_0,\lambda_1]$ the corresponding matrix
$T_j=T(\lambda)$ belongs to the case where $s>0$. Here $M=kl$, where
$2k$ is an order of $L$ and $l$ is its vector dimension.

Let $H^{as}$ be a direct sum of the Hamiltonian (Symplectic)
Spaces corresponding to the
different tails:
$$H^{as}=\oplus_{j=1}^{j=N} H_j^{2kl}$$
\noindent with natural skew-symmetric
 scalar-valued nondegenerate product defined
by the SWronskians in every tail $z_j$. For every solution $\psi$ for the
equation $L\psi=\lambda\psi$ on the whole graph $\Gamma$ we have its
{\bf asymptotic value}: 
$$\psi\rightarrow\psi^{as}\in H^{as}$$
\noindent where
$\psi^{as}_j\in H^{2kl}_j$ is this solution in the tail $z_j$.

\begin{th} The subspace $L^{as}\subset H^{as}$
of the asymptotic values for all solutions with given value of the
spectral parameter
$\lambda$, 
is a Lagrangian subspace of the half dimension equal to $Nkl$
(i.e. the SWronskian scalar product is identically equal to zero in it,
$<L^{as},L^{as}>=0$.)
\end{th}

Proof. This Theorem appeared the first time in the work \cite{N3} for the special
cases. The general proof is more or less the same as in  this special case.
Essentially the property of the asymptotic plane to be Lagrangian
 is a Topological Phenomenon, following directly from
the fact that the SWronskian is a cycle. For any pair of solutions for the
equation $L\psi=\lambda\psi, L\phi=\lambda\phi$ on the whole graph $\Gamma$
we have a cycle of the form:
$$W(\phi,\psi)=\sum_{j=1}^{j=N}a_jz_j+(finite)$$
\noindent where $z_j$ is a tail as a geometric cycle near infinity. Let me
remind that far enough in the tail
our operator is presented as one on the discretized line.
However, only differences can be extended to the cycles on the whole graph.
Therefore  we can express our SWronskian  through the differences only:
$$W(\psi,\phi)=\sum_{t=2}^{t=N}   b_t(z_1-z_t)+(finite)$$
\noindent Comparing these formulas, we see that
$$<\psi^{as},\phi^{as}>=\sum_j a_j=0$$
\noindent by the definition of the scalar product $<,>$ in the space $H^{as}$.

It is easy to see that  the plane $L^{as}$ of the asymptotic value of the solutions
extended to the whole graph $\Gamma$
is given by the number of equations equal to the half of the dimension
of the space $H^{as}$.
At the same time, we established the fact that this plane is Lagrangian
(i.e. the scalar product in it is equal to zero). The dimension of Lagrangian
plane is always less or equal to the half. Therefore it is equal to half exactly.
Our theorem is proved.

Let us point out that the complexified
asymptotic space $H^{as}$  in any noncritical
real point $\lambda$ has natural direct decomposition (with scalar
product of different parts equal to zero):
$$H^{as}=H_+\oplus H_-\oplus H_{bounded}$$
\noindent where the subspaces are defined in the following way:

Subspace  $H_-$ has dimension $(2p+q)N$. It contains all asymptotic
solutions with decay in every tail $z_j,j=1,\ldots,N$ for
 $n\rightarrow+\infty$;

 Subspace $H_+$ contains the solutions corresponding to the eigenvalues
 of the asymptotic monodromy matrix $T(\lambda)$ outside of the unit circle,
 $|\mu_|>1$; they are  increasing for $n\rightarrow\infty$ in every tail.
 The dimension of this subspace is also $(2p+q)N$;

 Subspace $H_{bounded}$ of the dimension $2sN$ corresponds to the unimodular
 eigenvalues; after complexification there is a natural decomposition
 $$H_{bounded}=H_{in}\oplus H_{out}$$
 \noindent on the waves $\psi_{j,in}$ and $\psi_{j,out}$
 coming inside and outside correspondingly in the tail $z_j$.
 It means precisely that the {\bf in}-part corresponds to the eigenvalues $\mu$
 of the monodromy matrix $T$
 with positive real parts and the {\bf out}-part is complex adjoint. 
 We have vector-functions
 $\psi_{j,in}^i=\bar{\psi}_{j,out}^i$ for the real $\lambda$,
 \noindent such that for the different indices $i$ they
  correspond to the different eigenvalues in the same tail $z_j$,
 and have zero symplectic scalar product,
 $$<\psi_{j,in}^i,\psi_{j,out}^t>=a(\lambda)\delta^{it}, a\neq 0$$

 \begin{deff} We call the interval on the real line {\bf
 generic and nonsingular}
 if the following requirements are satisfied:

  it does not
 contain critical points (i.e. the numbers $(p,q,s)$ are not changing
 in it, and all eigenvalues are nonmultiple);

  the intersection
 of the Lagrangian Plane $L^{as}(\lambda)$ with the subspace $H_-\oplus
 H_{bounded}$  has the dimension exactly equal to $sN$;

  the projection of this intersection
 on the subspace generated by the vectors $\psi_{j,in}^i$ for all $j,i$
 should be 'onto' (after complexification).
 \end{deff}

 For the  {\bf Generic Operators} the spectrum consists of such intervals
 separated by the isolated points which should be  passed transversally
 in the natural sense (see above their Jordan structure, but we require
 transversality for the interaction of the different tails also).
{\bf  Obviously, for the real big enough $|\lambda |$ we always have $s=0$.}

Therefore there is a finite number of finite intervals
 with nonzero values of $s>0$ only.

\begin{deff} Let the Scattering Matrix $S^{j',i}_{j,i'}(\lambda)$
for any generic nonsingular interval
on the $\lambda$-line be defined using the  complex
basis  of the  intersection
$$L^{as}(\lambda)\cap (H_-\oplus H_{bounded})$$
\noindent taken in the form
$$e_j^i=\psi_{j;in}^i+\sum_{j',i'}s^{j',i}_{j,i'}\psi^{i'}_{j';out}modulo
(H_-)$$

\end{deff}

 \begin{th}The Scattering Matrix $S$ defined above is a Unitary Symmetric
 Matrix for the real generic nonsingular values of $\lambda$.
 \end{th}

The proof of this theorem is parallel to the special case of  second
 order operators (see \cite{N4}).

{\it As it was written already in \cite{N3},
  it follows directly from the Lagrangian property of the plane
   $L^{as}\in H^{as}$.
   Take the basis $\psi_{in}+S\psi_{out}$ in the complexification
   of this plane for real $\lambda$. Different vectors of this basis
    have a zero scalar
   product with each other. This property implies that
   the matrix $S$ is symmetric $S^t=S$.
    From the reality we have $\bar{\psi}_{out}=
   \psi_{in}$ and
    $$\phi=\psi_{out}+\bar{S}\psi_{in}$$
    \noindent is complex adjoint to the
   previous basis.
   The basis
   $$\bar{S}^{-1}\phi=\psi_{in}+\bar{S}^{-1}\psi_{out}$$
   \noindent is coinside with the first one. Therefore we have
   $\bar{S}^{-1}=S$ and $S^t=S$. One may think that we took a real basis
   on the Lagrangian plane in the form
   $$A\psi_{in}+\bar{A}\psi_{out}, S=A^{-1}\bar{A}$$
   It follows from the Lagrangian property that $A$ can be taken as a
   unitary matrix  $A\in U_{klN}$. By unitarity, we have $\bar{A}^t=A^{-1}$
   and $S=BB^t,B=A^{-1}\in U_{Nkl}$. Multiplying the matrix $B$ from the right
   by the
   arbitrary real orthogonal matrix $B'=BO$, we see that
   $$B'(B')^t=BOO^tB^t=BB^t$$
   \noindent Therefore the Scattering Matrix $S$ depends on the Lagrangian
   Plane only. This plane
    may be identified with a point in the space $U/O$.}

   So, the proof
 is exactly  the same as in \cite{N4} for the
 Strongly Stable Case where $s=M, p=q=0$. For the general case with $s>0$
 we have
to use the fact that the SWroskians of any vector in the subspace
$H_-$ with themselves and with any vector from the subspace $H_{bounded}$ are
 identically equal to zero. It is completely obvious because any
  eigenfunction
 from the subspace $H_-$ is exponentially decreasing far enough in the tail.
 Therefore this additional term in the definition of the basis above
 for $S-matrix$ is completely negligible. Theorem is proved.

 \begin{rem}
 For the case $p+q>0$ we may meet a new type of singularities where
the projection of the
intersection of the Lagrangian Plane $L^{as}(\lambda)$ with 
subspace $H_{bounded}\oplus H_-$  into the space $H_{bounded}$ has a
rank smaller than $ks$ (here $k$ is a number of tails).
 This case corresponds
to the discrete spectrum drawn in the continuous one.
\end{rem}

\centerline{\bf Appendix: S.Novikov and A.Schwarz}
\vspace{0.2cm}
\centerline{\Large Nonlinear Discrete Systems on Graphs}
\vspace{0.2cm}
As  already mentioned in  paragraph 1, the Symplectic Geometry
of Discrete Second Order Lagrangian
Systems on the discretized line $R$ (i.e. on the lattice $Z$) was started
in  work \cite{V} (the pioneering work of Aubrey is quoted in \cite{V}
where  specific important example was investigated).

It was  explained
at the end of  paragraph 1 and in paragraph 2
how to extend this construction to the higher order
linear systems on the discretized line and on the general Graphs.
Let us discuss here {\bf Nonlinear Discrete Lagrangian Systems on Graphs}.
Consider as before any locally finite Graph $\Gamma$ presented as a
1-dimensional
simplicial complex without ends (i.e. any vertex belongs to at least two
edges). Suppose the following data are given:

Family  of manifolds $M^l_P$  numerated by the vertices
$P\in \Gamma$;

Family $X$ of  the sets $Q$ of vertices $P_j\in Q$ such that the maximal distance
$d_{\max}(P_i,P_j)$between the vertices in any set $Q$
 is equal to $D$; normally this family contains exactly
 all ''maximal''  sets of the perimeter $D$ containing all minimal paths between
 two vertices if the ends belong to  $Q$; it should not contain any minimal paths
 longer than $D$, and any minimal path in it should be extendable to the
 path of the length $D$;

 Family of  $C^{\infty}$-functions  (the {\bf  Density of Lagrangian})
 $$\Lambda_Q:\prod_{P_j\in Q} M_{P_j}\rightarrow R$$

\noindent  Using this data, we define an {\bf Action} for any function $\psi$
on the set of  vertices such that $\psi(P)\in M_P$:
$$S\{\psi(P)\}=\sum_{Q\in X} \Lambda_Q(\psi(P_{i_1},\ldots),P_i\in Q$$

\noindent For the infinite graphs this sum often does not exist, but
we define the {\bf Euler-Lagrange Equation} in the standard way:
$$\frac{\delta S}{\delta \psi(P)}= \frac{\partial S}{\partial \psi_P}=0$$

\noindent Therefore only the  sets $Q$ containing the point $P$
are involved in the calculation of the last
variational derivative (which is an ordinary partial derivative for
the discrete systems). We call the union of the sets $Q$ containing the vertex $P$
a {\bf Combinatorial Neighborhood $U_P$ of the Point $P$ of the order $D$}.

There are different possibilities here:

I.The equation above is sufficient to express the function $\psi(P)$
through the values $\psi(P_j)$ in the neighboring points $P_j\in U_P$.
 This situation
looks typical for  {\bf Elliptic-Type Problems} like the Dirichlet Boundary
Problem and so on. For example, if the manifolds $M^l_{P_j}$ are compact
for all vertices, we may take a minimum. We can do this also in many cases
 if all functions
$\Lambda_Q$ are nonnegative (or bounded from below).

II. The equation above is sufficient for the {\bf Nondegenerate expression}
of  $\psi_{P_j}$ in any point
$P_j$ on the boundary of the combinatorial neighborhood $U_Q$ through
other points in the combinatorial neighborhood $U_P$, where
$d(P,P_j)=D$. This situation we call {\bf Dynamical }.
 In some cases, beginning
from the property of the last type, we may define also the Hyperbolic Type.
Let me point out that the {\bf Dynamical situation was considered in 
paragraph 1 for the linear systems}: in this case 
the nondegeneracy of the Symplectic Form, generated by the SWronskian,
was proved.

\begin{th}Let the nonlinear Discrete Euler-Lagrange System   and its
solution $\psi(P)$ be given. Consider the linearized
self-adjoint operator $L$ near
 the solution $\psi$ and two solutions for the equation:
$$L(\delta \psi_a)=0,a=1,2$$
\noindent The SWronskian $W(\delta \psi_1,\delta\psi_2)$
defines a closed differential vector-valued 2-form SW
 with values in the space
$H_1^{open}(\Gamma,C)$, on the space of
solutions for the Nonlinear Discrete Euler-Lagrange System above.
For the discretized line this form is nondegenerate for the
nondegenerate Dynamical Type Systems.
\end{th}

For the second order translation invariant systems (see below)
on the discretized line our
theorem follows from  work \cite{V}. As A.Veselov pointed out to me,
for the higher order translationally invariant systems on the discretized line
thisw theorem also can be extracted from  \cite{V}-see the article in
Russian Math
Syrveys, pp 6-7. We shall publish full proof
of this theorem in 
separate paper.

\begin{deff} We call the Discrete Action $S$ and the Variational Problem above
{\bf The Second Order Translation Invariant Problem} in any Graph
if all manifolds $M_P^l$ are equal to the same manifold $M$,
 all sets $Q$ contain the same number
 of points equal to two $D=2$,
 and all functions $\Lambda_Q$ are equal to the same function $\Lambda(P_1,P_2)$
 of two variables (i.e. defined in $M\times M$).
 \end{deff}
\begin{rem} We can define the Translation Invariant Systems of any order for
the discretized line--lattice $Z$. In the case of  order four we can define them
for the locally homogeneous Graphs, where all vertices meet the same
number of edges equal to $m$. The function $\Lambda_Q$ for every set $Q$
has $m+1$ variables, i.e. it maps $M\times M\times \ldots M$ into $R$.
\end{rem}

\end{document}